\documentclass[a4paper,11pt]{article}
\pdfoutput=1 

\usepackage{jcappub} 

\usepackage[T1]{fontenc} 

\title{\boldmath  Gravitational Leptogenesis in Axion Inflation with SU(2) gauge field}

\author{Azadeh Maleknejad}

\affiliation{School of Physics, Institute for Research in Fundamental Sciences (IPM),\\ P. Code. 19538-33511, Tehran, Iran}
\emailAdd{azade@ipm.ir}

\newcommand{\be}{\begin{equation}}
\newcommand{\ee}{\end{equation}}
\newcommand{\bse}{\begin{subequations}}
\newcommand{\ese}{\end{subequations}}
\newcommand{\bea}{\begin{eqnarray}}
\newcommand{\eea}{\end{eqnarray}}
\newcommand{\ba}{\begin{array}}
\newcommand{\ea}{\end{array}}

\newcommand{\nn}{\nonumber}

\def\dre_g{\delta\rho_g}
\def\dpe_g{\delta P_g}
\def\dqe_g{\delta q_g}
\def\dre{\delta\rho}
\def\dpe{\delta P}
\def\dqe{\delta q}

\def\mH{\mathcal{H}}

\def\YM1{\frac{\dot\phi^2}{a^2}}
\def\YM2{\frac{g^2\phi^4}{a^4}}

\newcommand{\mpl}{M_{\rm pl}}

\def\treh{T_{\rm reh}}

\def\x{\tilde{\tau}}
\def\th{\tilde{h}}
\def\tg{\tilde{\gamma}}

\abstract{We present an intrinsic leptogenesis mechanism in models of axion inflation with a classical SU(2) gauge field. The gauge field is coupled to the axion with a Chern-Simons interaction and comprises a tiny fraction of the total energy, $\frac{\rho_{_{\rm YM}}}{\rho_{_{\rm tot}}}\lesssim\epsilon^2$. However, it has spin-2 fluctuations which breaks the parity and leads to the generation of chiral gravitational waves during inflation. By the gravitational anomaly in SM, it naturally creates a net lepton number density, sufficient to explain the matter asymmetry. We show that this mechanism can generate the observed value of baryon to photon number density in a natural range of parameters and yet has a small chiral tensor power spectrum on large scales.

}

\begin{document}

IPM/P-2016/016

\maketitle

\flushbottom


\section{Introduction}

To the best of our knowledge, the cosmos is highly matter dominated. Understanding the origin of this mysterious asymmetry in the Universe has been a major challenge of the modern cosmology and particle physics. In an expanding background (which leads to the departure from thermal equilibrium), the necessary and sufficient conditions to create a matter asymmetry from symmetric initial conditions are the violation of baryon number B, as well as C and CP violation \cite{Sakharov:1967dj}. The baryon-antibaryon asymmetry can be quantified by the baryon to photon ratio at the present time as 
$\eta_{_{\rm B}}=\frac{n_B}{n_{\gamma}}\simeq6\times10^{-10}$ \cite{Ade:2015xua}. First proposed by Fukugita and Yanagida in \cite{FY}, leptogenesis is a class of scenarios that associates the cosmic baryon asymmetry to an initial lepton asymmetry in the early Universe. In the standard approach of leptogenesis, the standard model of particle physics is extended by adding massive right-handed neutrinos which provide the source of CP violation in the model. The right-handed massive neutrinos then decay and generate the initial net lepton number, L \cite{Fong:2013wr, Canetti:2012vf}. The sphaleron process which would be thermally activated in temperatures $T\gtrsim 130 ~\rm{GeV}$ \cite{D'Onofrio:2014kta}, violates both B and L but conserved $B-L$. Therefore, the generated lepton number can be converted into the observed baryon asymmetry by the electro-weak sphaleron process.

 As an alternative approach, \textit{gravi-leptogenesis} is a class of inflationary leptogenesis scenarios in which the parity violating quantum fluctuations generated during inflation leads to a net lepton number \cite{Alexander:2004us}. When the gravitational field is taken into account in the computation of the lepton and baryon number anomalies, one finds that $B-L$ is not conserved. In particular, the gravitational anomaly of the lepton current $J_L^\mu$, in the standard model is the contribution of curvature to the lepton current divergence \cite{AlvarezGaume:1983ig}
\bea\label{G-anomaly-in}
\nabla_\mu J_L^\mu=\bigg(\frac{N_{\!_{l-r}}}{16\pi^2}\bigg)\tilde R R,
\eea\nonumber
where $N_{\!_{\rm l-r}}=N_{\rm l}-N_{\rm r}$ is the number of left/right-handed fermion
degrees of freedom and $\tilde RR$ is the Pontryagin density defined as
 $R\tilde R
\equiv\frac12\epsilon^{\lambda\mu\nu\xi}R_{\lambda\mu\rho\sigma}R_{\nu\xi}^{~~\rho\sigma}$.

Unlike chiral anomaly, such terms do not present in the baryonic sector since quarks have both chirality components. Thus, the gravitational anomaly in the presence of a non-vanishing $\tilde RR$ (due to chiral GWs) may serve as a $B-L$ production mechanism during inflation. One realization of this mechanism is studied in \cite{Alexander:2004us}. In this model, the inflation is driven by an axion field $\varphi$ and a \textit{modified gravity} interaction of the form $P(\varphi)R\tilde R$ provides the source of parity violation. In order for this model to generate sufficient lepton number, one needs $P(\varphi)\sim10^3$. Alternative axion driven inflationary baryogenesis scenarios based on using gauge fields in Einstein gravity has been proposed and studied in \cite{Alexander:2011hz, Noorbala:2012fh, Maleknejad:2014wsa}. Another interesting inflationary baryogenesis has been introduced in \cite{Hertzberg:2013jba} in which baryon asymmetry is generated via an Affleck-Dine like mechanism.

Recently in \cite{Maleknejad:2016qjz}, we studied the cosmic perturbations in a generic axion inflation model with a small SU(2) gauge field\footnote{Other inflationary models involving SU(2) gauge field has been introduced and studied in \cite{Maleknejad:2011sq} (\textit{gauge-flation}) and \cite{Adshead:2012qe} (\textit{chromo-natural}). In the gauge-flation model, the non-Abelian gauge field is the inflaton itself which after the end of inflation acts like a dark radiation and damps. The chromo-natural model consists of an axion field with a standard cosine potential coupled to an SU(2) gauge field through $\frac{\lambda\chi}{f}F\tilde F$. When $\frac{\lambda}{f}\sim\frac{\mathcal{O}(10^3)}{\mpl}$, this model leads to slow-roll inflationary background, without requiring super-Planckian $f$. Both of gauge-flation and chromo-natural inflation models have been disfavored by Planck data \cite{Dimastrogiovanni:2012ew, Adshead:2013nka}.}.
Our central finding was the existence of a parameter regime in which the gauge field can simultaneously generate a detectable chiral gravitational wave signal and has a negligible contribution to the scalar fluctuations, in agreement with the current CMB observations. Here, the inflaton field is the axion which is coupled to the gauge field through a Chern-Simons interaction, $\frac{\lambda}{f}\textmd{tr}(F^a_{\mu\nu}\tilde F_a^{\mu\nu})$ with $\frac{\lambda}{f}\sim\frac{\mathcal{O}(10)}{\mpl}$. Thanks to its SU(2) algebra, the gauge field has an isotropic and homogeneous field value in the background with an energy density $\rho_{_{\rm YM}}\lesssim\epsilon^2\rho_{\rm tot}$. The perturbed gauge field has a spin-2 fluctuation which linearly coupled to the primordial gravitational waves and explicitly breaks the parity between its left- and right-handed polarizations. As a result, our large scale tensor power spectrum is the standard vacuum fluctuations of the gravitational waves $P_{_{\rm vac}}\simeq2\big(\frac{H}{\pi\mpl}\big)^2$ plus an extra circularly polarized term which is coming from the interaction with the SU(2) gauge field, $P_{_{\chi}}\simeq\mathcal{G}^2_{_{+}}(\xi_{\psi})\big(\frac{\rho_{_{\rm YM}}}{\rho_{\rm tot}}\big)\big(\frac{H}{\pi\mpl}\big)^2$\footnote{For reviews on gravitational waves in models in which the (pseudoscalar) inflaton is coupled to U(1) gauge fields see \cite{Guzzetti:2016mkm} and \cite{Domcke:2016bkh}. Note that the U(1) gauge field quanta are mixed to the gravitational waves at the nonlinear level through the interaction $\delta A+\delta A\rightarrow\delta g$. That then generates chiral gravitational waves. Moreover, $\delta A$ is also coupled to the inflaton as well (e.g. $\delta A+\delta A\rightarrow\delta \varphi$) and generates large amounts of non-Gaussianity. In other words, the resulting sourced gravity wave signal is correlated to the large scale non-Gaussianity.
However, the mixing between the non-Abelian gauge field and perturbations in the scalar and tensor sectors are
at the linear order and coming from different fluctuations \cite{Maleknejad:2016qjz}.}. In principle, the chiral part of the power spectrum $P_{_{\chi}}$ can be of the same order as the unpolarized vacuum power spectrum and leads to non-vanishing parity odd CMB correlations.

In this paper, we consider the possibility of using the gravitational leptogenesis mechanism in the model of \cite{Maleknejad:2016qjz} to explain the observed baryon asymmetry in the Universe. Since our setup generates intrinsic chiral gravitational waves and hence a $R\tilde R\neq0$, it can serve as an inflationary leptogenesis mechanism. In order to be as model-independent as possible, we consider an arbitrary potential for the axion which is able to support the slow-roll inflation. One of the most popular and well-motivated axion models of inflation is \textit{monodromy inflation} \cite{Silverstein:2008sg,Flauger:2009ab, McAllister:2014mpa, Easther:2013kla, Flauger:2014ana}. This inflationary mechanism is a string theoretic construction which motivates a broad class of axion potentials of the form $V(\varphi)=\mu^{4-p}\varphi^p+\Lambda^4 e^{-c(\frac{\varphi}{\varphi_0})^{p_{\Lambda}}}\cos(\frac{\varphi_0}{f}(\frac{\varphi}{\varphi_0})^q+\theta_0)$. For an extensive review on the models of axion inflation and gauge fields in the physics of inflation see \cite{Pajer:2013fsa, Long:2016jvd} and \cite{Maleknejad:2012fw}. We find that for a typical and reasonable values of its parameter space, this mechanism can explain the observed matter asymmetry in the Universe. The final lepton number is proportional to the energy density of the gauge field during inflation, which provides the source of P violation. It is also proportional to the difference between the energy density of the left- and right-handed gravitational waves at short wavelengths. We show that the generated baryon to photon ratio and the super-horizon power spectrum of tensor modes are not directly related. Therefore, the model can generate a sizeable lepton number while it leads to a small chiral power spectrum at large scales\footnote{The effects of $SU(2)$ gauge fields have been also studied in the \textit{post-inflationary} scenarios of leptogenesis. Among these alternative models, we can refer to \cite{Kusenko:2014lra} in which the standard model Higgs boson is coupled to the electroweak gauge bosons as well as \cite{Kusenko:2014uta} which considers a general axion field coupled to the electroweak gauge bosons.
}.


This paper is organized as follows. In Sec. \ref{basic-setup}, we briefly review the axion inflation model with an SU(2) gauge field of ref. \cite{Maleknejad:2016qjz}. Sec. \ref{GWs} focuses on the analytical study of the gravitational waves in the presence of the SU(2) gauge field with a small VEV. In Sec. \ref{Lepton-production}, we compute the lepton number which through the gravitational anomaly in the standard model is generated during inflation. We conclude in section \ref{conclusion} and some technical details are presented in Appendices A and B.

\section{Axion inflation coupled to an SU(2) gauge field}\label{basic-setup}

Our setup is a generic single field axion inflation model with an SU(2) gauge field sector in Einstein gravity. This model has been studied recently in \cite{Maleknejad:2016qjz} and in this section, we only give a short review. Here and throughout, the reduced Planck mass is set to unity, unless otherwise specified. The Lagrangian density of the axion inflation is given as
\bea\label{action}
\mathcal{L}_{\textmd{inf}} =\frac{R}{2}-\frac12\partial_\mu\varphi\partial^\mu\varphi-V(\varphi),
\eea
where $\varphi$ is the axion field with an arbitrary potential $V(\varphi)$, flat enough to support the slow-roll inflation. 
The axion field is coupled to an SU(2) gauge field, $A^a_{\mu}$, through the Chern-Simons interaction. 
The gauge field strength tensor is
\bea
F^a_{~\mu\nu}=\partial_\mu A^a_{~\nu}-\partial_\nu A^a_{~\mu}-g\epsilon^a_{~bc}A^a_{~\mu}A^b_{~\nu},
\eea
where $g$ is the gauge coupling and $a, b, c...$ are the indices of the $su(2)$ algebra. The gauge field theory of our setup is  
\bea\label{action}
\mathcal{L}_{A}=-\frac{1}{4}\bigg(F^a_{\mu\nu}F_a^{\mu\nu}+\frac{\lambda}{f}\varphi\ F^a_{\mu\nu}\tilde{F}_a^{\mu\nu}\bigg)\,,
\eea
where $\tilde{F}^{a\mu\nu}=\frac12\epsilon^{\mu\nu\lambda\sigma}F^a_{\lambda\sigma}$, $\lambda$ is a dimensionless parameter and $f$ is the axion decay constant. More inflationary models involving gauge fields has been reviewed in \cite{Maleknejad:2012fw}.

In the flat FRW background,
\bea
\label{FLRW}
ds^2=-dt^2+a(t)^2\delta_{ij}dx^{i}dx^{j},
\eea
and after fixing the gauge as $A^a_0=0$, we have an isotropic and homogeneous solution for the spatial part of the gauge field\footnote{
The isotropic and homogeneous configuration of the gauge field in the temporal gauge can be written in the geometrical form 
\bea\label{geometry-ansatz}
\quad A^a_{~\mu}(t)= \psi(t)e^a_{~\mu},
\eea%
in which $\{e^{\alpha}_{~\mu}\}$ are tetrads of FRW metric, $e^{0}_{~\mu}=n_{\mu}$ and $e^{a}_{~\mu}=a(t)\delta^a_{\mu}$,
where $n^{\mu}=(1,0,0,0)$ is the 4-velocity of the comoving observer \cite{Maleknejad:2011sq, Maleknejad:2016qjz}.}
\bea\label{ansatz}
\quad A^a_{~i}(t)= a\psi(t)\delta^a_{~i}\,,
\eea%
where $\psi$ is basically the effective field value of the gauge field in the background. The background configuration \eqref{ansatz} generates a $F_{\mu\nu}$ with electric and magnetic components as
\bea
E^a_i=-(H\psi+\dot{\psi})\delta^a_i \quad \textmd{and} \quad B^a_i=-g\psi^2\delta^a_i.
\eea

The total energy density $\rho_{_{\rm tot}}$, is 
\bea
\rho_{_{\rm tot}}=\frac12\dot{\varphi}^2+V(\varphi)+\rho_{_{\rm YM}},
\eea
where $\rho_{_{\rm YM}}$ is the energy density of the gauge field 
\bea
\rho_{_{\rm YM}}=\frac12\bigg(\vec{E}^a.\vec{E}_a+\vec{B}^a.\vec{B}_a\bigg).
\eea
For the purpose of this work, we are interested in the regime that the gauge field has a negligible effect in the background evolution. More precisely, the energy density of the gauge field is a small fraction of the total energy density
\bea
\frac{\rho_{_{\rm YM}}}{\rho_{_{\rm tot}}}\lesssim\epsilon^2.
\eea

The field equations of $\varphi$ and $\psi$ are respectively 
\bea\label{c-n.e.o.m}
&\ddot\varphi+3H\dot\varphi+V_{\varphi}=-3\frac{\lambda g}{f} \psi^2(\dot\psi+H\psi)\,,\\\label{eq-psi}
&\ddot\psi+3H\dot\psi+(2H^2+\dot H)\psi+2g^2\psi^3=\frac{\lambda g}{f}\psi^2\dot\varphi\,.
\eea
It is noteworthy to mention that the Chern-Simons interaction in the RHS of the above equations with a $\dot{\varphi}\neq0$, breaks the conformal symmetry of the gauge field and prevents its decay during inflation.

We are interested in the slow-roll inflation in which $\epsilon$ and $\eta$ 
\bea\label{SL-H}
\epsilon=-\frac{\dot H}{H^2},\quad  \eta=-\frac{\ddot H}{2H\dot H},
\eea
are very small. Moreover, the slow varying evolution of the gauge field can be quantified in terms of $\epsilon_{\psi}$ and $\eta_{\psi}$ as
\bea\label{vartheta}
\epsilon_{\psi}\equiv \frac{\dot\psi}{H\psi} \quad \textmd{and}  \quad \eta_{\psi}\equiv-\frac{\ddot{\psi}}{H\dot{\psi}}\,.
\eea
For later convenience, at this point, we define two dimensionless parameters 
\bea
\xi_{\psi}\equiv\frac{g\psi}{H} \quad \textmd{and} \quad \xi\equiv\frac{\lambda\dot{\varphi}}{2fH},
\eea
which during the slow-roll inflation are related as
\bea\label{xi-eq}
\xi\simeq\frac{(1+\xi_{\psi}^2)}{\xi_{\psi}}.
\eea

Demanding slow-varying evolution of the gauge field, \eqref{eq-psi} requires that $\xi\sim1$ and $\lambda/f\sim1/\sqrt{\epsilon}$.
The energy density of the gauge field is almost constant during slow-roll inflation. As the axion rolls down its potential, however, the Chern-Simons interaction gradually injects some of the axion's energy to the gauge field and increases $\rho_{_{\rm YM}}$. At the end of inflation, $\dot{\varphi}$ starts oscillating around the minimum of the potential and the gauge field acts like a self-interacting (dark) radiation sector, $\rho_{_{\rm YM}}\propto1/a^4$.

\section{Gravitational Waves and the Gauge Field}\label{GWs}

The fluctuations of the SU(2) gauge field contributes to the cosmic perturbations and leads to new theoretical and observational features.
As far as our current discussion and the gravitational anomaly is concerned, we only need to know the tensor fluctuations. Therefore, in this section, we focus on the (symmetric, traceless and divergence-free) tensor perturbations. The perturbed metric is 
\bea\label{metric-pert}%
\delta\!_{_{\rm T}}g_{\mu\nu}=a^2\gamma_{ij},
\eea
where $\gamma_{ij}$ is the gravitational wave and $\delta\!_{_{\rm T}}$ denotes the tensor sector of the fluctuations. Perturbing the SU(2) gauge field around its isotropic configuration \eqref{ansatz}, we have another tensor fluctuation, $\tg_{ij}$, given as\footnote{The second term in $\delta\!_{_{\rm T}} A^a_{~i}$ is the induced space-time transformations on the gauge field. More precisely, one can write \eqref{gauge-field-pert} as $\delta\!_{_{\rm T}} A^a_{~i}=\delta^{aj}a\tg_{ij}+\psi\delta\!_{_{\rm T}} e^a_{i}$, where $\{\delta e^a_\mu\}$ are the perturbed tetrads. (See \cite{Maleknejad:2016qjz} for more geometrical details.) }
\bea \label{gauge-field-pert}
\delta\!_{_{\rm T}} A^a_{~i}=\delta^{aj}a\big(\tg_{ij}+\frac{\psi}{2}\gamma_{ij}\big).
\eea
As a result, the Yang-Mills term in the action generates an anisotropic inertia in the linear order energy-momentum tensor
\bea\label{pi_T}
\pi^T_{ij}\simeq2H\psi\bigg((\xi^2_{\psi}-1)H\tg_{ij}-\dot{\tg}_{ij}
+\frac{\xi_{\psi}}{a}\partial_k\big(\epsilon^{kl}_{~~(i}\tg_{j)l}\big)\bigg),
\eea
which modifies the field equation of the gravitational wave as
\bea \label{T-gf}
\ddot \gamma_{ij}+3H \dot \gamma_{ij}-\frac{1}{a^2}\nabla^2\gamma_{ij}=2\pi^T_{ij}.
\eea
We emphasis that in order to have a linear order anisotropic inertia in \eqref{pi_T}, the gauge fields should be turned on at the background level ($\psi\neq0$).
The quadratic action and field equation of $\tg_{ij}$ are given in Appendix A. Going to the Fourier space, we can diagonalize the system of \eqref{T-gf} and \eqref{tg-ij} in terms of circular polarization modes.

We can expand $\gamma_{ij}$ and $\tg_{ij}$ in terms of the right- and left-handed polarization states 
\bse
\begin{align}
&\gamma_{ij}(\tau,\textbf{x})=\frac{1}{\sqrt{2}a}\sum_{\sigma=R,L}\int \frac{d^3k}{(2\pi)^\frac32} h_{\sigma}(\tau,\textbf{k})e^{\sigma}_{ij}(\textbf{k})e^{i\textbf{k}.\textbf{x}},\\ &\tg_{ij}(\tau,\textbf{x})=\frac{1}{2\sqrt{2}a}\sum_{\sigma=R,L}\int \frac{d^3k}{(2\pi)^\frac32} \th_{\sigma}(\tau,\textbf{k})e^{\sigma}_{ij}(\textbf{k})e^{i\textbf{k}.\textbf{x}},
\end{align}
\ese
where $\{\frac{h_{\sigma}}{\sqrt{k}},\frac{\tilde h_{\sigma}}{\sqrt{k}}\}$ are the canonically normalized fields and $e^{R,L}_{ij}$ are the circular polarization tensors.
For a wave vector $\textbf{k}=(0,0,k)$, the right- and left-handed modes are defined as $h_{_{R,L}}\equiv a(\gamma_{11}\pm i\gamma_{12})/2$.
In the Fourier space, it is useful to work in terms of two new variables
\be\label{T-def}
\x\equiv-k\tau \quad \textmd{and} \quad \tilde\mH\equiv\frac{\mH}{k},
\ee
where $\tau$ is the conformal time and $\mH=aH$. Upon using the slow-roll inflation, $\mH\simeq-(1+\epsilon)/\tau$, we can read $\x\simeq\frac{k_{\rm phy}}{H}$ where $k_{\rm phy}=k/a$ is the physical momentum.

\subsection{Field equations of tensor modes}

In the Fourier space and in terms of the canonically normalized fields $h_{\sigma}$ and $\th_{\sigma}$, the field equation \eqref{T-gf} can be read as
\bea\label{eq-h--}
\partial^2_{\x}h_{_{R,L}}+\bigg(1-\big(2-\epsilon\big)\tilde{\mathcal{H}}^2\bigg)h_{_{R,L}}\simeq S^T_{_{R,L}}(\th_{_{R,L}}),
\eea
where $S^T_{_{R,L}}(\th_{_{R,L}})$ is the linear source term of the gravitational waves, given in \eqref{pi_T}
\bea
S^T_{_{R,L}}(\th_{_{R,L}})\simeq2\psi\tilde\mH\bigg(\partial_{\x}\th_{_{R,L}}+(\xi_{\psi}^2\tilde{\mH}\mp\xi_{\psi})\th_{_{R,L}}\bigg).
\eea
The field equation of $\tilde h_{_{R,L}}$ is 
\bea\label{v-eq-app}
\partial^2_{\x}\th_{_{R,L}}(\textbf{k},\tau)+\bigg(1\mp\frac{2(\xi+\xi_{\psi})}{\x}+\frac{2\xi\xi_{\psi}}{\x^2}\bigg)\tilde{h}_{_{R,L}}(\textbf{k},\tau)\simeq0,
\eea
where we neglect the sub-dominant RHS of \eqref{tg-ij}.
The solution of equation \eqref{eq-h--} can be decomposed in terms of the homogeneous and the particular solutions, as
\bea\label{hR}
h_{_{R,L}}(\textbf{k},\x)=h^{G}_{_{R,L}}(\textbf{k},\x)+h^{S}_{_{R,L}}(\textbf{k},\x),
\eea
where $h^{G}$ is the solution of the homogeneous equation (coming from vacuum fluctuations). 
The homogeneous solution is unpolarized and therefore is given by a single function $h(\x)$
\bea\label{h-hankel}
h(\x)\simeq-\sqrt{\frac{\pi\x}{2}}H^{^{(1)}}_{\nu_T}(\x) \quad \textmd{for} \quad \nu_T\simeq\frac32+\epsilon.
\eea
The particular solution of the gravitational wave is given by the Green's integral below
\bea\label{h-sigma}
h^{\!^{s}}_{_{R,L}}(\x)=\int_{\x}^{\infty}G(\x,\x')S^T_{_{R,L}}(\x')d\x',
\eea
where $G(\x,\x')$ is the retarded Green's function of equation \eqref{eq-h--}
\bea
\label{Green}
G(\x,\x')\simeq\bigg(\frac{\x'-\x}{\x'\x}\cos(\x'-\x)-(1+\frac{1}{\x\x'})\sin(\x'-\x)\bigg)\Theta(\x'-\x),
\eea
where $\Theta(\x-\x')$ is the Heaviside step function .

The most general form of the solutions of \eqref{v-eq-app} can be written in terms of $M_{\kappa_{\rm \sigma},\mu}(-2i\x)$ and $W_{\kappa_{\rm \sigma},\mu}(-2i\x)$ Whittaker functions. After imposing the Banch-Davis vacuum condition for $\frac{1}{\sqrt{k}}\th_{_{R,L}}(\x)$, we obtain  $\th_{_{\rm R,L}}$ as\footnote{ Note that only $W_{\kappa,\mu}(-2i\x)$ function represents the positive frequency solution in the Minkowski limit, $\x\gg1$.} 
\bea\label{tilde-h}
\th_{\sigma}(\x)=e^{i\kappa_{\sigma}\pi/2} W_{\kappa\!_{\sigma},\mu}(-2i\x),
\eea
where 
\bea\label{re-def}
\kappa_{_{R,L}}=\mp i\big(\xi+\xi_{\psi}\big) \quad \textmd{and} \quad \mu^2=\frac14-2\xi\xi_{\psi}.
\eea

In the next subsection, we determine the analytical form of the super-horizon power spectrum. In subsection \ref{WKB}, using the WKB approximation, we work out the explicit form of the chiral gravitational waves in the short wavelength scales.

\subsection{Long wavelength Power spectrum}

\begin{figure}[h!]
\begin{center}
\includegraphics[width=0.65\textwidth]{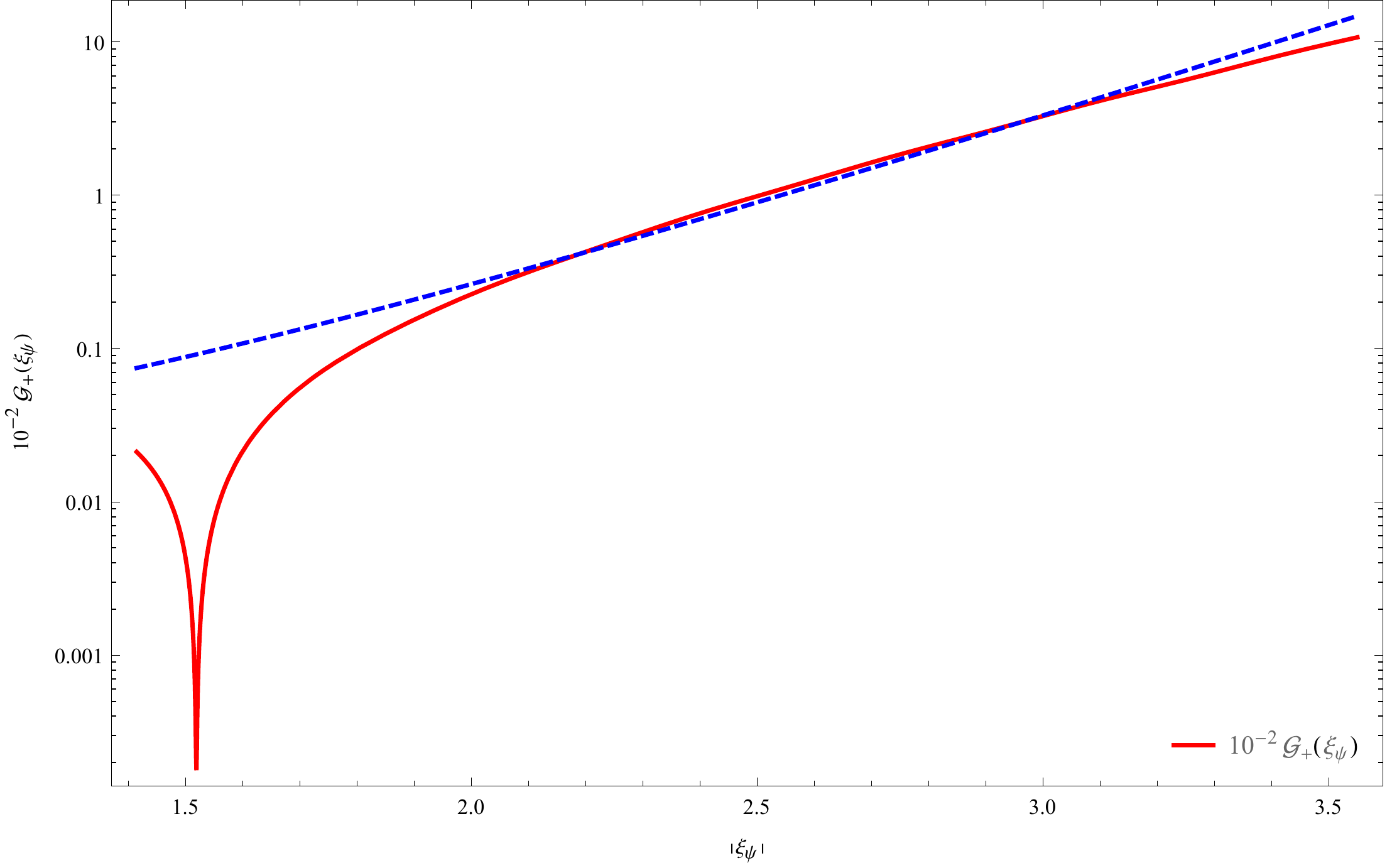}
\end{center}
\caption{The pre-factor $\mathcal{G}_{_{+}}(|\xi_{\psi}|)$ with respect to $|\xi_{\psi}|$. Since $h^s_{_{+}}\propto\big(\frac{\rho_{\rm YM}}{\rho_{\rm tot}}\big)^{\frac12}\mathcal{G}_{_{+}}$ where $\big(\frac{\rho_{\rm YM}}{\rho_{\rm tot}}\big)^{\frac12}\lesssim10^{-2}$, we rescaled $\mathcal{G}_{_{+}}$. The (blue) dotted line is $\frac{10^{-2}}{50}e^{\frac{\pi}{2\xi_{\psi}}(1+2\xi_{\psi}^2)}$ which provides a good approximation for $10^{-2}\mathcal{G}_{_{+}}$ in the $2<|\xi_{\psi}|<3$.}\label{IR-Gamma}   
\end{figure}

Doing the Green's integral \eqref{h-sigma} in the limit that $\x\ll1$, we obtain the particular solution of the gravitational waves as \cite{Maleknejad:2016qjz}
\bea\label{super-hs}
\gamma^{\!^{s}}_{_{+}}(\tau,k)\simeq \mathcal{G}_{_{+}}(\xi_{\psi})\bigg(\frac{\bar{\rho}_{_{\rm YM}}}{\bar{\rho}_{_{\rm tot}}}\bigg)^{\!\frac12}\bigg(\frac{H}{k^{\frac32}}\bigg) \quad  \textmd{and} \quad \gamma^{\!^{s}}_{_{-}}(\tau,k)\simeq  0.
\eea
The prefactor $\mathcal{G}_{_{+}}(\xi_{\psi})$ is given as 
\bea\label{Int-expres}
\mathcal{G}_{_{+}}(\xi_{\psi})\!&\simeq&\!e^{\frac{i\pi}{2}\kappa_{_{\!+}}}\frac{2\sqrt{(1+\xi^2_{\psi})}}{\xi^2_{\psi}} \bigg(
\frac{(i\xi_{\psi}+1)\Gamma(-\kappa_{_{\!+}})}{\Gamma(\frac12-\kappa_{_{\!+}}-\mu)\Gamma(\frac12-\kappa_{_{\!+}}+\mu)}+\frac{(i\xi_{\psi}-1)}{\Gamma(1-\kappa_{_{\!+}})}\bigg)\Gamma(\frac12-\mu)\Gamma(\frac12+\mu),\nonumber
\eea 
here $\kappa_{+}=-\frac{i(1+2\xi_\psi^2)}{|\xi_{\psi}|}$. In case of $\psi>0$ ($\psi<0$), the plus polarization is the right-(left-) handed polarization mode. In figure \ref{IR-Gamma}, we present $\mathcal{G}_{_{+}}(\xi_{\psi})$ with respect to $|\xi_{\psi}|$.
As we see, in the regime $2<\xi_{\psi}<3$, we can approximate $\mathcal{G}_{_{+}}(\xi_{\psi})$ by $\frac{1}{50}e^{\frac{\pi}{2\xi_{\psi}}(1+2\xi_{\psi}^2)}$. For $\xi_{\psi}>3$, $\mathcal{G}_{_{+}}(\xi_{\psi})\gtrsim 5$ and leads to a large r which is disfavoured by the latest joint analysis of Planck and BICEP2/Keck array measurements \cite{Ade:2015tva}. Therefore, we are interested in the regime $\xi_{\psi}\lesssim3$.

The power spectrum of the particular solution of gravitational waves is given as
\bea\label{plz-power}
P_{\gamma^{s}_{+}}=\frac{8\pi k^3}{(2\pi)^3}|\gamma^s_{+}|^2\simeq \mathcal{G}^2_{_{+}}(\xi_{\psi}) \bigg(\frac{\bar{\rho}_{_{\rm YM}}}{\bar{\rho}_{_{\rm tot}}}\bigg)   \bigg(\frac{H}{\mpl\pi}\bigg)^2 \quad\quad  \textmd{and} \quad P_{\gamma^{s}_{-}}(\tau,k)\simeq  0,
\eea
which is circularly polarized and is related to the chirality parameter of CMB, $\chi\equiv\frac{P_{_{R}}-P_{_{L}}}{P_{_{vac}}}$, as
\bea
\chi(\xi_{\psi})=\frac{sP_{\gamma^{s}_{+}}}{P_{_{vac}}}=s\mathcal{G}^2_{_{+}}(\xi_{\psi})\bigg(\frac{\bar{\rho}_{_{\rm YM}}}{\bar{\rho}_{_{\rm tot}}}\bigg), \quad \textmd{where} \quad
s=\textmd{sign}(\psi).
\eea
Finally, given the fact that $\gamma^{G}$ and $\gamma^{S}$ are uncorrelated and working out \eqref{h-hankel}, we obtain the power spectrum of gravitational waves as
\bea\label{power-T}
P_{_{T}}\simeq \bigg(2+|\chi(\xi_{\psi})|\bigg)\big(\frac{H}{\pi\mpl}\big)^2,
\eea
where the second term is the contribution of the gauge field's tensor fluctuations to the super-horizon power spectrum.

\subsection{Chiral fluctuations in the short wavelength scales}\label{WKB}
In the deep inside horizon limit, $\x\gg 1$, we can use WKB analysis in \eqref{v-eq-app} to determine $\th_{\sigma}$ as
\bea
\th_{_{R,L}}(\x)\simeq\frac{1/\sqrt{2}}{\sqrt[4]{1\mp2(\xi+\xi_{\psi})/\x}}\exp\bigg(\mp i(\xi+\xi_{\psi})\ln(\x)\bigg)\exp(i\x).
\eea
Upon using the above solution and the WKB approximation in \eqref{h-sigma}, we obtain the particular solution of $h_{\sigma}$ as
\bea\label{sub-h}
h^{\!^{s}}_{_{R,L}}(\x)\simeq-\frac{(\xi_{\psi}\mp i)\psi}{\sqrt{2}(\xi+\xi_{\psi})}\exp\bigg(\mp i(\xi+\xi_{\psi})\ln(\x)\bigg)\exp(i\x).
\eea
We stress that the canonically normalized gravitational wave is $\frac{1}{\sqrt{k}}h^{\!^{s}}_{_{R,L}}(\x)$.
As we see in \eqref{sub-h}, the two polarization states are different even in short wavelength limit. That can be seen more explicitly in the energy density of each polarization state of gravitational waves. The energy-momentum tensor of the gravitational waves is given as
\bea
T^{^{\rm GW}}_{\mu\nu}=\frac{\mpl^2}{4} \langle\gamma_{ij,\mu}\gamma^{ij}_{~~,\nu}\rangle.
\eea
The energy density of each polarization state can be read as
\bea\label{gw-energy}
\rho^{\rm GW}_{\sigma}=\int \rho_{\sigma,k}(\tau)d\ln{k},
\eea
in terms of the spectral energy density which is
\bea
\rho_{\sigma,k}(\tau)=\frac{\mpl^2k^3}{8\pi^2 a^2} \langle\bigg(\frac{h^{*}_{\sigma}(\tau,k)}{a}\bigg)'\bigg(\frac{h_{\sigma}(\tau,k)}{a}\bigg)'\rangle.
\eea
Both $h^{G}_{_{\rm R,L}}$ and $h^{S}_{_{\rm R,L}}$ contributes to the energy density. The vacuum gravitational waves are unpolarized. However, the particular solution is chiral and its right- and left-handed polarizations have different energy densities. Using \eqref{sub-h} in \eqref{gw-energy} and doing the integral up to a cut-off scale $k_{\rm phy}=\Lambda$, we have
\bea\label{density-GW}
\Omega^{\rm GW}_{L-R}=\frac{1/72\pi^2}{(\xi_{\psi}+\xi)}\bigg(\frac{\bar{\rho}_{_{\rm YM}}}{\mpl^4}\bigg)\bigg(\frac{\Lambda}{H}\bigg)^4,
\eea
where $\Omega^{\rm GW}_{L-R}$ is the difference between the density parameters of left- and right-handed polarizations, where $\Omega^{\rm GW}_{\sigma}\equiv\frac{\rho^{\rm GW}_{\sigma}}{3\mpl^2H^2}$. Using the slow-roll relation \eqref{xi-eq}, we can simplify that as 
\bea\label{density-GW-1}
\Omega^{\rm GW}_{L-R}=\frac{\xi_{\psi}/72\pi^2}{(1+2\xi_{\psi}^2)}\bigg(\frac{\bar{\rho}_{_{\rm YM}}}{\mpl^4}\bigg)\bigg(\frac{\Lambda}{H}\bigg)^4.
\eea

\section{Gravitational anomaly and baryon asymmetry}\label{Lepton-production}

Considering the gravitational interactions in the computation of the lepton and baryon number anomalies, one finds that $B-L$ is not conserved. From the gravitational anomaly of the lepton current $J_L^\mu$, in the standard model \cite{AlvarezGaume:1983ig}, we have
\bea\label{G-anomaly}
\nabla_\mu J_L^\mu=\bigg(\frac{N_{\!_{l-r}}}{16\pi^2}\bigg)\tilde R R,
\eea
where $N_{\!_{\rm l-r}}=N_{\rm l}-N_{\rm r}$ is the number of left/ right-handed fermion
degrees of freedom and $\tilde RR$ is the Pontryagin density defined as
 $$\tilde R
R\equiv\frac12\epsilon^{\lambda\mu\nu\xi}R_{\lambda\mu\rho\sigma}R_{\nu\xi}^{~~\rho\sigma}.$$
In order for this mechanism to work, we need both of $R\tilde R$ and $N_{\!_{\rm l-r}}$ to be non-zero. In the standard model of particle physics in which we can integrate out massive right-handed neutrinos, $N_{\!_{\rm l-r}}\lvert_{SM}=3$. However, at energy scales higher than the mass scale of massive right-handed neutrinos, predicted by see-saw mechanism, they are propagative and $N_{\!_{\rm l-r}}\leqslant1$ \cite{Minkowski:1977sc}. Considering the homogeneous and isotropic FRW background metric, $\tilde{R}R$
vanishes in the background, while the tensor fluctuations of the perturbed metric \eqref{metric-pert} contributes as
\bea\label{RR-hij}
\tilde R R=-\frac{2}{a^4}\epsilon^{ijk}\big(\gamma''_{jl}\partial_i \gamma'_{lk}-\partial_m
\gamma'_{jl}\partial^2_{im}\gamma_{lk}+\partial_l \gamma'_{jm}\partial^2_{mi}\gamma_{kl}\big),
\eea
where prime denotes a derivative with respect to the conformal time. The Pontryagin density is a parity odd quantity which can be non-zero in the presence of chiral gravitational waves. Recalling that our model naturally generates chiral gravitational waves, this mechanism can produce lepton asymmetry during inflation. Relying on the thermally activated electroweak instantons (sphalerons), that generated lepton number could finally be transformed to baryon asymmetry. In this section, we study the possibility of using this mechanism as a leptogenesis scenario to explain the observed baryon asymmetry in the Universe.

\subsection{Primordial lepton production rate and number density}

 Using \eqref{T-def} in \eqref{RR-hij}, and after some lengthy calculations which is presented in Appendix \ref{3rd-App}, we have the quantum expectation value of the Pontryagin density as
\bea\label{RR-sec}
\langle\frac{N_{\!_{\rm l-r}}}{16\pi^2}\tilde{R}R(\tau)\rangle=\frac{1/8\pi^4}{a^4}\sum_{\sigma=R,L}\!\lambda_{\sigma}\int k^2dk N_{\!_{l-r}}
\frac{d}{d\tau}\bigg(\bigg|\bigg(\!\gamma^{\!s}_{\sigma}(\tau,k)\!\bigg)'\bigg|^2-k^2\bigg|\gamma^{\!s}_{\sigma}(\tau,k)\bigg|^2\bigg),
\eea
where $\lambda_{_{R,L}}=\pm1$. Using the fact that $\gamma^{\!s}_{\sigma}(\tau,k)=\sqrt{2}\big(\frac{H}{k}\big)\x h^{\!s}_{\sigma}(\x)$ and $\x=\frac{k_{\rm phy}}{H}$ is the normalized physical momentum, we obtain
\bea\label{RR-hRL-}
\langle\frac{N_{\!_{ \rm l-r}}}{16\pi^2}\tilde{R}R(\tau)\rangle=-\frac{H^6}{4\pi^4}\sum_{\sigma=R,L}\!\lambda_{\sigma}\!\int^{\frac{\Lambda}{H}}_0\!N_{\!_{\rm l-r}}\!(\x)
\frac{d}{d\x}\bigg(\bigg|\bigg( \x h^{\!s}_{\sigma}(\x)\bigg)_{\!\!\x}\bigg|^2 -\bigg|\x h^{\!s}_{\sigma}(\x)\bigg|^2\bigg)\x^3d\x.
\eea
Here we considered a see-saw theory with three massive \textit{right-handed neutrinos} where the mass of the heaviest one, $M_{\rm R}\approx \Lambda$. In fact, in sufficiently low energy scales, we can integrate out all the right handed neutrinos and $N_{\!_{\rm l-r}}\lvert_{SM}=3$ while $N_{\!_{\rm l-r}}$ vanishes for $k_{\rm phy}\gtrsim\Lambda$. For simplicity, we consider the following approximation for $N_{\!_{\rm l-r}}$ with respect to the energy scale
\bea\label{nl_R}
N_{\!_{\rm l-r}}(k_{\rm phy})=\left\{
\begin{array}{ll} 3\, \quad \quad  &k_{\rm phy}<\Lambda
\\    0\, \quad \quad &k_{\rm phy}\geq\Lambda.
\end{array}\right.
\eea%
Since $\Lambda\gg H$, it implies that here we consider three massive right-handed neutrinos with mass scales higher than $H$.
Note that only the chiral part of the gravitational waves contributes to the integrand of \eqref{RR-hRL-}.
Using \eqref{sub-h} and \eqref{nl_R} in \eqref{RR-hRL-}, we obtain 
\bea
\langle\frac{N_{\!_{\rm l-r}}}{16\pi^2}\tilde RR(\tau)\rangle=\frac{\Lambda^4/4\pi^4}{(\xi_{\psi}+\xi)}\bigg(\frac{\bar{\rho}_{_{YM}}}{M^4_{\rm pl}}\bigg),
\eea
which as we expect, is proportional to the energy density of the gauge field at the background.

Working out RHS of \eqref{G-anomaly} for the lepton current density $J^{\mu}_{_{L}}=(n_{_{L}},\vec{J}_{_{L}})$, we obtain the following equation for the physical lepton number density\footnote{In \eqref{eq-B-L}, we used the fact that the spatial average of $\partial_iJ^i_{_{L}}$ is negligible, corresponding to assumption that the surface integral $\int_{\!\partial_i V}\!J^i_{L}$ vanishes.}, $n_{_{L}}$, as
\bea\label{eq-B-L}
\dot{n}_{_{L}}+3Hn_{_{L}}=\Gamma\!_{_{L}}(\tau),
\eea
where the lepton production rate $\Gamma\!_{_{L}}$ is 
\bea\label{rate}
\Gamma\!_{_{L}}(\tau)=\frac{1}{4\pi^4}\frac{\Lambda^4}{(\xi_{\psi}+\xi)}\frac{\bar{\rho}_{_{YM}}}{M^4_{\rm pl}}.
\eea
Integrating \eqref{eq-B-L}, we obtain the physical lepton number density at a time $\tau$ as
\bea
n_{_{L}}(\tau)=\frac{1}{a^3(\tau)}\int^{\tau}_{-\infty}d\tau'a^4(\tau')\Gamma\!_{_{L}}(\tau').
\eea
For the exact de Sitter limit in which $\bar\rho_{_{\rm YM}}\propto a^{-4}$ and $(\xi_{\psi}+\xi)\propto a^{-1}$, the number density is damping like $a^{-3}$. However, for an inflationary slow-roll background, $\Gamma_{_{L}}$ has a slow varying evolution and up to the first order in slow-roll, we have $n_{_{L}}(\tau)\simeq\frac{\Gamma\!_{_{L}}(\tau)}{3H(\tau)}$. Upon using \eqref{rate} and the slow-roll relation \eqref{xi-eq}, we obtain
\bea\label{Lepton-number}
n_{_{L}}(\tau)\simeq\frac{\xi_{\psi}/12\pi^4}{(2\xi_{\psi}^2+1)}\frac{\bar{\rho}_{_{YM}}}{M^4_{\rm pl}}\big(\frac{\Lambda}{H}\big)^4 H^3,
\eea
 which slowly increases with time. The generated lepton number is related to the difference between the energy density of left- and right-handed polarizations of gravity waves at the energy scale $\Lambda$ ( eq. \eqref{density-GW})
\bea
n_{_{L}}\simeq\frac{6}{\pi^2}\Omega^{\rm GW}_{L-R}(\Lambda)H^3.
\eea
The generated lepton number has the following noteworthy features.\\
$\bullet$ As one may expect, $n_{_{L}}$ is proportional to the energy density of the gauge field during inflation which provides the source of parity violation.\\
$\bullet$ The factor $H^3$ is the inverse of the volume (horizon) size during inflation, and has the same unit as $n$.\\
$\bullet$ The lepton number is proportional to $\Lambda^4$ where $\Lambda$ is roughly the mass of the heaviest right-handed neutrino. Above that energy scale, our three right-handed neutrinos are dynamical and $N_{\!_{\rm l-r}}=0$. More precisely, $n_{\rm L}$ is proportional to the difference between the energy density of left- and right-handed polarizations at the energy scale $k_{\rm phy}=\Lambda$. Given the fact that $\tilde RR$ becomes negligible after inflation, we expect that $n_{\rm L}$ scales as $a^{-3}$ afterward, \textit{i.e.}
\bea\label{nL-scale}
a^3(\tau)n_{\rm L}(\tau)=a_{\rm inf}^3n_{\rm L, \rm inf},
\eea
where $\tau$ is a time after the end of inflation. Therefore, the lepton number density at the end of reheating would be
\bea\label{nL-reh}
n_{_{L},\rm reh}\simeq \frac{\xi_{\psi}/12\pi^4}{(2\xi_{\psi}^2+1)}\frac{\bar{\rho}_{_{YM}}}{M^4_{\rm pl}}\big(\frac{\Lambda}{H}\big)^4 H^3\bigg(\frac{a_{\rm inf}}{a_{\rm reh}}\bigg)^3,
\eea
where $a_{\rm reh}$ is the scale factor at the end of reheating.

\subsection{Baryon to photon number density}

In order to connect the generated lepton number to $\eta_{_{\rm B}}$, we need to determine the number density of photons at the present time.
 The energy density at reheating is given by
\bea\label{reheat}
\rho_{\rm reh}(\treh)=\frac{\pi^2}{30}g_{\rm eff} \treh^4\,,
\eea
where $g_{\rm eff}=427/4$ is the number of relativistic degrees of freedom at the time of reheating and
$\treh$ is the reheating temperature. 
The photon number density at the time of reheating is
\bea\label{photon}
n_{\gamma,\rm reh}=\frac{2\zeta(3)}{\pi^2}T^3_{\rm reh},
\eea
where $\zeta(x)$ is the Riemann zeta function and $\zeta(3)=1.2$.
Here, we consider the phenomenological reheating model below
\bea\label{reh-inf}
\rho_{\rm reh}=\sigma \bigg(\frac{a_{\rm inf}}{a_{\rm reh}}\bigg)^{\!4}\rho_{\rm inf}   ,
\eea
in which $\sigma$ is the \textit{efficiency of the reheating} process and relates $\rho_{\rm reh}$ and the energy density at the end of inflation, $\rho_{\rm inf}$. The reheating temperature is then given as
\bea\label{Treh}
\bigg(\frac{T_{\rm reh}}{\mpl}\bigg)=\sqrt{3}\bigg(\frac{ H}{\mpl}\bigg)^{\frac12}\bigg(\frac{\sigma}{g_{\rm eff}}\bigg)^{\frac14}\bigg(\frac{a_{\rm inf}}{a_{\rm reh}}\bigg).
\eea
One can also characterize the dynamics of reheating in terms of the inflaton decay rate $\Gamma_{\varphi}$, which is related to the reheating temperature as
\be
T_{\rm reh}\simeq\bigg(\frac{90}{g_{\rm reh}\pi^2}\bigg)^{\frac14}\sqrt{\Gamma_{\varphi}\mpl}.
\ee

Combining \eqref{reheat}-\eqref{reh-inf}, we can read the number density of photons during reheating as 
\bea\label{ngamma-reh}
n_{\gamma, \rm reh}=\frac{6\sqrt{3}\zeta(3)}{\pi^2}  \bigg(\frac{\sigma}{g_{\rm eff}}\bigg)^{\frac34}\big(\frac{a_{\rm inf}}{a_{\rm reh}}\big)^3
(H\mpl)^{\frac32}\,.
\eea
One can also write $n_{\gamma, \rm reh}$ in terms of the inflaton decay rate as
\bea\label{ngamma-reh-g}
n_{\gamma, \rm reh}=\frac{2\zeta(3)}{\pi^2}  \bigg(\frac{90}{g_{\rm eff}\pi^2}\bigg)^{\frac34}
(\Gamma_{\varphi}\mpl)^{\frac32}\,.
\eea

 Relying on the electroweak sphaleron processes, the generated primordial matter-antimatter
asymmetry in the lepton sector can transform to the baryonic sector \cite{KRS,FY}. The final baryon asymmetry is given as follows
\bea\label{eta-}
\frac{n_{B0}}{n_{\gamma0}}=c_{\rm sph}\frac{g_{\rm eff,0}}{g_{\rm eff}}\bigg(\frac{n_{L,\rm reh}}{n_{\gamma,\rm reh}}\bigg),
\eea
where $c_{\rm sph}=28/79$ is the sphaleron conversion factor in the standard model and $g_{\rm eff,0}=43/11$. Hence, from the combination of \eqref{nL-reh}, \eqref{ngamma-reh} and \eqref{eta-}, we obtain the desired $\eta_{_{\rm B}}$ as 
\bea\label{n/s}
\eta_{_{\rm B}}\simeq 3\times 10^{-4}\frac{\xi_{\psi}/\sigma^{\frac34}}{(2\xi_{\psi}^2+1)}\frac{\bar{\rho}_{_{YM}}}{M^4_{\rm pl}}\big(\frac{\Lambda}{H}\big)^4   \bigg(\frac{H}{\mpl}\bigg)^{\frac32}.
\eea
We stress that $H$ and $\bar{\rho}_{_{\rm YM}}$ are the values of the Hubble parameter and energy density of the gauge field during inflation.
 As one may expect, $\eta_{_{\rm B}}$ is inversely related to the efficiency of the reheating process. In terms of the inflaton decay rate, the above result reads as
 \bea\label{n/ng-g}
\eta_{_{\rm B}}\simeq 8.5\times 10^{-4}  \bigg(\frac{\mpl}{\Gamma_{\varphi}}\bigg)^{\frac32}\frac{\xi_{\psi}}{(2\xi_{\psi}^2+1)}\frac{\bar{\rho}_{_{YM}}}{\bar{\rho}_{\rm inf}}\big(\frac{\Lambda}{H}\big)^4 \bigg(\frac{a_{\rm inf}}{a_{\rm reh}}\bigg)^3  \bigg(\frac{H}{\mpl}\bigg)^{5}.
\eea

\subsection{Comparison with observations}\label{observation}

Here we compare the theoretical prediction of our model for the baryon to photon ratio in \eqref{n/s} with the observed value $\eta_{_{\rm B}}\simeq6\times 10^{-10}$ \cite{Ade:2015xua}. It was showed in \cite{Maleknejad:2016qjz} that a successful inflation in agreement with the CMB data requires that $\psi\lesssim10^{-2}$, $\sqrt{2}\lesssim\xi_{\psi}<3$ and for a GUT scale inflation with $H\sim10^{-6}\mpl$ and $f\simeq0.1\mpl$, we have $g\sim 10^{-4}$ and $\lambda\sim1$. In this mechanism, a successful leptogenesis then requires 
\bea\label{--n/s}
\frac{\xi_{\psi}/\sigma^{\frac34}}{(2\xi_{\psi}^2+1)}\bigg(\frac{\bar{\rho}_{_{\rm YM}}}{\bar{\rho}_{_{\rm inf}}}\bigg)\bigg(\frac{\Lambda}{H}\bigg)^4 \bigg(\frac{H}{ M_{\rm pl}}\bigg)^\frac72\simeq\frac23\times10^{-6},
\eea
which for a typical value $\xi_{\psi}\sim 1$ and recalling that $\frac{\bar{\rho}_{_{\rm YM}}}{\bar{\rho}_{_{\rm inf}}}\lesssim10^{-4}$, we arrive at
\bea\label{--n/s}
\bigg(\frac{\sigma}{10^{4}}\bigg)^{3/4}\lesssim\bigg(\frac{\Lambda}{H}\bigg)^4 \bigg(\frac{H}{ M_{\rm pl}}\bigg)^\frac72.
\eea
That condition can be fulfilled for typical and reasonable values of reheating temperature
and the energy scale $\Lambda$. For instance, considering inflation at GUT-scales, $H\sim 10^{-6}\mpl$, for $\Lambda\sim 10-100H$, we need a reheating efficiency $\sigma \lesssim10^{-13}-10^{-18}$. Recalling that $\frac{\Gamma_{\varphi}}{\mpl}\sim\sigma^{\frac12}(\frac{H}{\mpl})^\frac{7}{3}$, that then gives an upper bound for the inflaton decay rate as $\Gamma_{\varphi}\lesssim10^{-20}\mpl$.

Associating $\Lambda$ with the mass of the right-handed neutrinos, it implies that at least one of our right-handed neutrinos would have a mass as high as $M_{\rm R}\sim10^{-5}\mpl$.

Within the supersymmetric extension of SM, gravitinos production gives an upper bound on the reheating temperature $\treh<10^4$ TeV
\cite{T-reh}. On the other hand, since the sphaleron process is thermally activated in temperatures $T\gtrsim 130 ~ \rm{GeV}$, this mechanism requires a reheat temperature $\treh\gtrsim130$ GeV. From \eqref{Treh}, we have the following relation for the reheating temperature 
\bea
\bigg(\frac{T_{\rm reh}}{\mpl}\bigg)\simeq 10^{-4}\sigma^{\frac14}\frac{a_{\rm inf}}{a_{\rm reh}}\,,
\eea
which corresponds to a reheat temperature $\treh\lesssim 10^{10}$ GeV.

\section{Conclusions}\label{conclusion}

In this work, we present a natural inflationary leptogenesis mechanism in models of axion inflation with a small SU(2) gauge field. In this scenario, the Chern-Simons interaction of the gauge field with the axion (where $\dot{\varphi}\neq0$) provides the source of the CP violation during inflation and leads to a non-vanishing $R\tilde R$. With the gravitational anomaly of the lepton number current in the standard model, this setup generates a net lepton number during inflation. The cosmic perturbations of this setup have been recently studied in \cite{Maleknejad:2016qjz}. The gauge field has a small VEV and negligible contribution to the inflationary background. However, its quantum fluctuation has a spin-2 sector which is coupled to the gravitational waves and modifies its evolution. Therefore, in addition to their standard vacuum fluctuations, the gravitational waves have a circularly polarized sector as well. The fact that this new component is polarized makes them distinguishable from the unpolarized vacuum fluctuations. That changes the gravitational waves in both the short and long wavelength scales. In the short scales, the chiral sector leads to a difference between the energy density of the left- and right-handed polarization states. In particular, the difference between the left- and right-handed energy parameters up to a cut-off scale $\Lambda$ is $\Omega^{\rm GW}_{L-R}=\frac{\xi_{\psi}/72\pi^2}{(1+2\xi_{\psi}^2)}\big(\frac{\bar{\rho}_{_{\rm YM}}}{\mpl^4}\big)\big(\frac{\Lambda}{H}\big)^4$. In the long wavelength scales, the tensor power spectrum has a chirality factor, given as $\chi(\xi_{\psi})\simeq\mathcal{G}^2_{_{+}}(\xi_{\psi})\big(\frac{\bar{\rho}_{_{\rm YM}}}{\bar{\rho}_{_{\rm tot}}}\big)$. That modifies the super-horizon power spectrum of the gravitational waves as $P_{_{T}}\simeq \big(2+\chi(\xi_{\psi})\big)\big(\frac{H}{\pi\mpl}\big)^2$. In fact, the parity odd tensor power is the unique observational feature of this leptogenesis mechanism which makes it distinguishable from the standard thermal leptogenesis scenarios. In the latter scenario, the source of parity violation is provided by the decay of massive neutrinos and after inflation. However, in the former case, due to the interactions with the gauge field the parity violation happens during inflation. 
In particular, the chiral gravitational waves lead to the existence of parity odd correlations in CMB, e.g. $\langle TB\rangle\neq0$ and $\langle EB\rangle\neq0$ which are zero in the standard scenarios.

The net lepton number which is generated in this setup is $n_{_{L}}\propto\frac{\bar{\rho}_{_{YM}}}{M^4_{\rm pl}}\big(\frac{\Lambda}{H}\big)^4 H^3$. As we expect, that is proportional to the energy density of the gauge field during inflation, which provides the source of P violation. The factor $H^3$ is the inverse of the horizon size during inflation.
Our lepton number is proportional to $\Lambda^4$ where $\Lambda$ is roughly the mass of the heaviest right-handed neutrino $ M_{\nu_R}\approx\Lambda$. More precisely, this setup is based on assuming that standard model neutrinos are Majorana and a type I seesaw mechanism is responsible for neutrino masses. For simplicity, we consider three massive right-handed neutrinos with equal masses which have been integrated out and as a result, we arrive at the $R\tilde R$ as the effective interaction. Since the right-handed neutrinos are not active in this minimal scenario, the standard thermal leptogenesis does not contribute to the final lepton asymmetry.  
The generated lepton number is related to the difference between the energy density of left- and right-handed polarizations of GW as $n_{_{L}}\simeq\frac{6}{\pi^2}H^3\Omega^{\rm GW}_{L-R}(\Lambda)$.
 Relying on the electroweak sphaleron processes, this generated matter asymmetry in the lepton sector can eventually transform to the baryonic sector. It is noteworthy to mention that in this setup inflation ends in a
(dark) radiation dominated Universe which may have interesting features for the (pre)reheating era. Moreover, the interaction $\varphi F\tilde F$ provides a natural decay channel for the inflaton during (pre)reheating. For the purpose of this work, we consider a phenomenological reheating model and the details of the reheating is beyond the scope of this paper.  Considering a phenomenological reheating model with the efficiency parameter $\sigma$, we determined the present time photon number density and $\eta_{_{\rm B}}$. This model predicts a baryon to photon ratio equal to $\eta_{_{\rm B}}\simeq3\sigma^{-3/4}(\frac{H}{\pi\mpl})^{\frac32}\Omega^{\rm GW}_{L-R}(\Lambda)$.

This mechanism can explain the observed matter asymmetry in the Universe for a typical and reasonable values of its parameter space. In particular,  
for a GUT scale inflation with $H_{\rm inf}\sim 10^{-6}\mpl$ and $\Lambda\sim (10-100)H_{\rm inf}$, we need a reheating efficiency $\sigma\gtrsim10^{-13}-10^{-18}$.  That corresponds to an inflaton decay rate $\Gamma_{\varphi}\lesssim10^{-20}\mpl$ and a reheat temperature $\treh\lesssim 10^{10}$ GeV. Associating $\Lambda$ with the mass of the right-handed neutrinos, it implies that, in order for this mechanism to work, at least one of our right-handed neutrinos should be as massive as $M_{\rm R}\sim 10 H_{\rm inf}$. For simplicity here, however, we considered three massive right-handed neutrinos with the same mass scales. The large scale tensor power spectrum and in particular $\chi(\xi_{\psi})$ are much more sensitive to the value of $\xi_{\psi}$ than the generated $\eta$. As a result, our  model can generate sufficient lepton number and at the same time a small chiral power spectrum at large scales, in the regime  $\sqrt{2}\lesssim\xi_{\psi}\lesssim3$ which for a GUT scale inflation and $f\sim 0.1\mpl$ corresponds to a $g\sim 10^{-4}$ and $\lambda\sim1$.

\section*{\small Acknowledgment}

It is a pleasure to thank Paolo Creminelli and Raphael Flauger for fruitful discussion and their encouragements. I am grateful to the hospitality of Stanford University where this  work has been initiated and the Galileo Galilei Institute for theoretical physics (GGI) and ICTP during its completion. I acknowledge support from Allameh Tabatabaii grant of Boniad Melli Nokhbegan Iran.

\appendix

\section{Quadratic action of $\tg_{ij}$}

In order to make this work self-sufficient, here we present the quadratic action of $\tg_{ij}$ and work out its linear field equation. The tensor fluctuations of the gauge field generates a strength tensor which at the linear order has the form
\bea\label{Fmunu-tensor}%
\delta\!_{_{\rm T}} F^a_{~0i}&=&\delta^{aj}\big(a\tg_{ij}+\frac{a\psi}{2}\gamma_{ij}\dot{\big)}\,,\\
\delta\!_{_{\rm T}}
F^a_{~ij}&=&2\big(a\delta^{ak}\partial_{[i}\tg_{j]k}-a^2g\psi\epsilon^{ak}_{~~~[j}\tg_{i]k
}\big)+\psi\big(a\delta^{ak}\partial_{[i}\gamma_{j]k}-a^2g\psi\epsilon^{ak}_{~~~[j}\gamma_{i]k
}\big)\,.
\eea
The spin-2 fluctuations of the gauge field contribute to the second order action and leads to a quadratic action for $\tg_{ij}$
\bea
\label{2ndts}
&&\delta\!_{_{2}}S_{\tg}\simeq\frac12\int d^3x dt a^3\biggl(
\big(\dot{\tg}_{ij}\big)^2-\big(\frac{\partial_k\tg_{ij}}{a}\big)^2-2\xi\xi_{\psi}H^2\tg_{ij}^2-2(\xi
+\xi_{\psi})H\epsilon^{ijk}\tg_{kl}\frac{\partial_i\tg_{jl}}{a}\nn\\
&&~~~~~~+2H\psi\big(\dot{\gamma}_{ij}+\xi\epsilon^{ilk}\frac{\partial_k\gamma_{jl}}{a}\big)\tg_{ij}
\biggr).
\eea
From the action, we obtain the field equation of $\tg_{ij}$ as
\bea\label{tg-ij}
\ddot\tg_{ij}+3H\dot\tg_{ij}-\frac{1}{a^2}\nabla^2\tg_{ij}+\frac{2(\xi+\xi_{\psi})}{a}H\partial_l\big(\epsilon^{lk}_{~~(i}\tg_{j)k}\big)+\xi\xi_{\psi}H^2\tg_{ij}=H\psi\bigg(\dot{\gamma}_{ij}+\xi\epsilon^{ikl}\frac{\partial_k\gamma_{jk}}{a}\bigg).~~~~~
\eea
As we see, there are two parity odd terms in the above field equation. This system can be diagonalized in terms of the Circular polarization states and the P violating terms have different signs for the right- and left-handed modes. 

\section{Pontryagin density of circularly polarized gravitational waves}\label{3rd-App}

In this appendix, we determine the explicit form of the Pontryagin density in terms of the right- and left-handed circular polarizations of the gravitational waves. The Pontryagin density (also known as the Chern-Pontryagin term) is a parity violating term defined as
\bea\label{RtR}
\tilde R
R\equiv\frac12\epsilon^{\lambda\mu\nu\xi}R_{\lambda\mu\rho\sigma}R_{\nu\xi}^{~~\rho\sigma},
\eea
where $\epsilon^{\lambda\mu\nu\xi}$ is the totally antisymmetric tensor and
$R^{\mu}_{~\nu\lambda\sigma}$ is the Riemann tensor. That can be express as the divergence of the Chern-Simons topological current
\bea
K^{\mu}=\epsilon^{\mu\nu\lambda\sigma}\Gamma^{\beta}_{\nu\alpha}\bigg(\partial_{\lambda}\Gamma^\alpha_{\sigma\beta}+\frac23\Gamma^\alpha_{\lambda\gamma}\Gamma^\gamma_{\sigma\beta}\bigg),
\eea
where $\{\Gamma^{\mu}_{\nu\lambda}\}$ is the Christoffel connection.
 Using the perturbed metric around the FRW background in \eqref{metric-pert}, the tensor fluctuations of the metric, $\gamma_{ij}(\tau,\textbf{x})$, contributes to the second order $\tilde R R$
\bea
\tilde R R=-\frac{2}{a^4}\epsilon^{ijk}\big(\gamma''_{jl}\partial_i \gamma'_{lk}-\partial_m
\gamma'_{jl}\partial^2_{im}\gamma_{lk}+\partial_l \gamma'_{jm}\partial^2_{mi}\gamma_{kl}\big),
\eea
where prime denotes a derivative with respect to the conformal time ($d\tau=a^{-1}dt$).

We can write $\tilde{R}R$ in terms of the Fourier components of the gravitational waves 
$\gamma_{ij}(\tau,\textbf{k})$, as
\bea\label{RR-app1}
\tilde R R(\tau,\textbf{x})=-\frac{2i\epsilon^{ijk}}{a^4} \iint
\frac{d^3kd^3k'}{(2\pi)^{3}} k'^i \bigg(\gamma''_{jl}(\tau,\textbf{k})
\gamma'_{lk}(\tau,\textbf{k}')+\textbf{k}.\textbf{k}'\gamma'_{jl}(\tau,\textbf{k})
\gamma_{lk}(\tau,\textbf{k}')\bigg)e^{i(\textbf{k}+\textbf{k}').\textbf{x}}+\mathcal{D
},\nonumber\\
\eea
where $\mathcal{D}$ is a total derivative term and therefore vanishes.
In terms of right- and left-handed polarizations
\bea
\gamma_{ij}(\tau,\textbf{k})=\sum_{\sigma=R,L}\gamma_{\sigma}(\tau,\textbf{k})e^{\sigma}_{ij}(\tau,\textbf{k}),
\eea
and after neglecting the total derivative term, we can mostly simplify $\tilde RR$ as\footnote{It is noteworthy to mention that if one naively write
\eqref{RR-app1} in terms of $\gamma_{R,L}$, the result would be
\bea
\tilde R R(\tau,\textbf{x})=-\frac{8i}{a^4} \iint \frac{d^3kd^3k'}{(2\pi)^{3}} k'
\big(\gamma''_{R}(\tau,\textbf{k})
\gamma'_{L}(\tau,\textbf{k}')+\textbf{k}.\textbf{k}'\gamma'_{R}(\tau,\textbf{k})\gamma_{L}(\tau,\textbf{k}')-R\leftrightarrow
L\big)e^{i(\textbf{k}+\textbf{k}').\textbf{x}},\nonumber
\eea
which is not a Hermitian operator. In fact, in writing the last term, one has to not only exchange R and L ($R\leftrightarrow L$), but also change the order of operators.} 
\bea\label{A-RR}
\tilde R R(\tau,\textbf{x})=-\frac{8i}{a^4} \iint \frac{d^3kd^3k'}{(2\pi)^{3}} k'
\big(\gamma''_{R}(\tau,\textbf{k})\gamma'_{L}(\tau,\textbf{k}')+\textbf{k}.\textbf{k}'\gamma'_{R}(\tau,\textbf{k})\gamma_{L}(\tau,\textbf{k}')-c.c.\big)e^{i(\textbf{k}+\textbf{k}').\textbf{x}}.~~~~~
\eea
Recalling \eqref{hR}, we can express the gravitational wave in terms of two uncorrelated terms
\bea\label{hR-app}
\gamma_{_{R,L}}(\textbf{k},\x)=\gamma^{G}_{_{R,L}}(\textbf{k},\x)+\gamma^{S}_{_{R,L}}(\textbf{k},\x),
\eea
where $\gamma^{G}$ is coming from the vacuum fluctuations and $\gamma^{S}$ is sourced by the gauge field. One can expand $\gamma^{G}(\textbf{k},\x)$ and $\gamma^{S}(\textbf{k},\x)$ as
\bse\label{hG-SS}
\begin{align}
\gamma^G_{R}(\tau,\textbf{k})=\frac{1}{\sqrt{k}}\bigg(\hat{a}_{R,\textbf{k}}\gamma(\tau,\textbf{k})+\hat{a}^{\dag}_{L,-\textbf{k}}\gamma^{*}(\tau,-\textbf{k})\bigg),\\
\gamma^S_{R}(\tau,\textbf{k})=\frac{1}{\sqrt{k}}\bigg(\hat{b}_{R,\textbf{k}}\gamma^{\!^{s}}_{R}(\tau,\textbf{k})+\hat{b}^{\dag}_{L,-\textbf{k}}\gamma_{L}^{\!^{s}*}(\tau,-\textbf{k})\bigg).
\end{align}
\ese
where the creation and annihilation operators $\hat{a}_{\textbf{k}}$ and
$\hat{b}_{\textbf{k}}$, satisfy the standard canonical relations $\big($\textit{i.e.}
$[\hat{a}^{\sigma}_{\textbf{k}},\hat{a}^{\sigma'\dagger}_{\textbf{k}'}]=\delta^{\sigma\sigma'}\delta^{(3)}(\textbf{k}-\textbf{k}
')\big).$ By definition, the left-handed polarization is given as $h_{L}(\tau,\textbf{k})=h^{*}_{R}(\tau,-\textbf{k})$.

Upon using \eqref{hR-app} and \eqref{hG-SS} in \eqref{A-RR} as well as the fact that $\gamma^{G}$ is unpolarized and uncorrelated with $\gamma^{S}$, we find the expectation value of $\tilde{R}R$ as
\bea\label{RRR}
\langle\tilde{R}R(\tau)\rangle=\frac{4}{a^4}\sum_{\sigma=R,L}\!\lambda_{\sigma}\int \frac{d^3k}{(2\pi)^{3}}
\frac{d}{d\tau}\bigg(\bigg|\bigg(\!\gamma^{\!s}_{\sigma}(\tau,\textbf{k})\!\bigg)'\bigg|^2-k^2\bigg|\gamma^{\!s}_{\sigma}(\tau,\textbf{k})\bigg|^2\bigg)  \quad \textmd{with} \quad \lambda_{_{R,L}}=\pm1,\nonumber
\eea
which after assuming the statistical isotropy of the primordial fluctuations, gives 
\bea\label{RRR-r}
\langle\tilde{R}R(\tau)\rangle=\frac{2/\pi^2}{a^4}\sum_{\sigma=R,L}\!\lambda_{\sigma}\int k^2dk
\frac{d}{d\tau}\bigg(\bigg|\bigg(\!\gamma^{\!s}_{\sigma}(\tau,k)\!\bigg)'\bigg|^2-k^2\bigg|\gamma^{\!s}_{\sigma}(\tau,k)\bigg|^2\bigg),
\eea
where $\lambda_{_{R,L}}=\pm1$.
As a result, only the (chiral) particular part of the gravitational waves contributes to the Pontryagin density term.

\end{document}